\documentclass[11pt,a4paper]{article}

\usepackage[no-natbib-sort]{my-jheppub}

\usepackage{graphicx}
\usepackage[english]{babel}
\usepackage{amsthm}

\usepackage{lipsum}
\usepackage{xargs}  
\usepackage[pdftex,dvipsnames]{xcolor} 
\usepackage[colorinlistoftodos,prependcaption,textsize=tiny]{todonotes}
\newcommandx{\unsure}[2][1=]{\todo[linecolor=OliveGreen,backgroundcolor=OliveGreen!25,bordercolor=OliveGreen,#1]{#2}}
\newcommandx{\correct}[2][1=]{\todo[linecolor=red,backgroundcolor=red!25,bordercolor=red,#1]{#2}}
\newcommandx{\improvement}[2][1=]{\todo[linecolor=blue,backgroundcolor=blue!25,bordercolor=blue,#1]{#2}}
\newcommandx{\tonote}[2][1=]{\todo[linecolor=yellow,backgroundcolor=yellow!25,bordercolor=yellow,#1]{#2}}

\usepackage[nodayofweek]{date time}

\def \ed  {\end{document} }

\begin{document}

\date{\currenttime}


\title{A Note on (Non)-Locality \\
in Holographic Higher Spin Theories}

\emailAdd{d.ponomarev@imperial.ac.uk}

\author{Dmitry Ponomarev}

\affiliation{Theoretical physics group, Blackett Laboratory, Imperial College London,   SW7 2AZ, U.K.}


\abstract{It was argued recently that the holographic higher spin theory features non-local interactions.
We further elaborate on these results using the Mellin representation. The main difficulty previously encountered on this way is that the Mellin amplitude for the free theory correlator is ill-defined.
To resolve this problem, instead of literally applying the standard definition, we propose to define this amplitude by linearity
using decompositions, where each term has the associated Mellin amplitude well-defined.
Up to a sign, the resulting amplitude  is equal to  the Mellin amplitude for the singular part of the quartic vertex in the bulk theory
and, hence, can be used to analyze bulk locality.
From this analysis we find that the scalar quartic self-interaction vertex in the holographic higher spin theory has a singularity of a  special form,
which can be distinguished from generic bulk exchanges. We briefly discuss the physical interpretation of such singularities and  their relation to the Noether procedure.}

\unitlength = 1mm

\today
\date {}
\begin{flushright}\small{Imperial-TP-DP-2017-{02}}\end{flushright}

\maketitle

\def \lR {L}

\def \ll {{\cal \ell}}

\def \be {\begin{equation}}\def \ee {\end{equation}}

\def \ads {AdS$_4$\ }
\def \iffa {\iffalse} 

\def \ed {\bibliography{KHSA}{}
\bibliographystyle{utphys} \end{document}}

\section{Introduction}
\label{AppC}
There is an overwhelming evidence that in conventional sense local higher spin theories do not exist in flat space.
This evidence comes from numerous no-go results obtained within a wide range of approaches 
\cite{Weinberg:1964ew,Coleman:1967ad,Aragone:1979hx,Metsaev:1991thesis,Bekaert:2010hp,Dempster:2012vw,Joung:2013nma}.
This problem was recently reconsidered using the light-cone \cite{PonSkv:2016} and the manifestly covariant \cite{Taronna:2017wbx,Roiban:2017iqg}
approaches.
At the same time,  recent analysis \cite{Ponomarev:2016lrm,Ponomarev:2017nrr}
 indicates that at least in the self-dual sector there exist  consistent local 
higher spin theories with  properties very similar to those of their lower spin counterparts. One option could be to stop 
here and declare that higher spin theories in flat space cannot go beyond the self-dual sector. Alternatively, one can try to relax locality
in some controllable way and push on with parity invariant higher spin theories.

So far in constructing higher spin theories locality was the main guiding principle and relaxing it will make the problem
too ill-defined, see, for example, \cite{Barnich:1993vg}\footnote{Analogous statements can also be proven within the light-cone 
deformation procedure \cite{Ponomarev:2016cwi}.}. Also, usually non-locality has some undesirable physical consequences,
such as superluminal propagation, Ostrogradsky instability, etc. This implies that the requirement of locality should be replaced
with another guiding principle that would ensure both that the problem of higher spin interactions is well-posed
and that physical pathologies are absent. 

We do not have much to say on what these new guiding principles should be. However, what we can do instead is to look at 
 higher spin theory in AdS and see how locality is violated there. This can give us a better idea of what locality violations to expect
 for putative parity invariant higher spin theories in flat space.
The advantage of considering the AdS setting is that in this case the higher spin theory is known, at least at the level of scattering
amplitudes. Indeed, via holography in the simplest setup  these can be identified with the correlators of the
free $O(N)$ vector model  \cite{Sezgin:2002rt,Klebanov:2002ja}\footnote{Note that the idea 
of computing higher spin scattering amplitudes in AdS space from ``the S-matrix'' of singletons was
discussed long 
before higher spin holography acquired its modern form  \cite{Flato:1980zk,Fronsdal:1986ui}.}. The latter, in turn, can be easily computed.
Thus, by studying the free theory correlators one can learn whether the bulk higher spin theory is local and if not, how exactly locality
is violated. Another attractive feature of the holographic higher spin theory is that, being dual to a healthy theory on the 
boundary, it is unlikely to suffer from any serious problems.

Proceeding along these lines, recently  it was argued \cite{Sleight:2017pcz} that holographic higher spin theory is non-local, see
\cite{Bekaert:2015tva,Taronna:2016ats,Bekaert:2016ezc} for earlier discussions. This conclusion is based on the fact that
via combining conformal block decompositions of bulk and boundary four-point functions in different channels
 one can show that the conformal block decomposition for the bulk theory contact interaction contains single trace conformal blocks.
 This, in turn, can be regarded as an indication that the quartic interaction vertex in the holographic higher spin theory is non-local.

While this argument, indeed, strongly suggests that the relevant vertex is non-local, it would be important to have an
explicit formula clearly characterizing this non-locality. An explicit formula for the 
non-local part of the contact interaction may help us to understand whether the associated non-locality is 
general enough to trivialize the Noether procedure\footnote{In the higher spin literature the procedure of perturbative construction of a gauge
invariant action, see e.g. \cite{Berends:1984rq}, is called the Noether procedure.} along the lines of the argument \cite{Barnich:1993vg}.
This information may be then used to amend the standard Noether procedure in a way, that makes
the problem of higher spin interactions a well-defined one. The resulting approach or, rather its flat space version, can then be employed to construct higher spin theories in the Minkowski space. 

In this paper we analyze the Mellin amplitude for the four-point correlator in the free $O(N)$
vector model.  Previously, this question  was addressed in \cite{Taronna:2016ats,Bekaert:2016ezc}. These attempts encountered problems that result
into an ill-defined Mellin amplitude. 
Here we propose to \emph{define} this amplitude using the superposition principle. To be more precise, one can present the boundary correlator in the form of a superposition of interfering processes and then define its Mellin amplitude as the sum of Mellin amplitudes for each individual process. We then use this Mellin amplitude and the fact that the boundary correlator up to a sign is equal to the singularity of the contact four-point interaction in the bulk higher spin theory to characterize non-locality of the latter.

This paper is organized as follows. In section \ref{loc} we review how locality is defined in AdS. Then, in 
section \ref{lochs} we discuss how locality can be tested for holographic higher spin theories using the conformal
block decomposition. In particular, we review \cite{Sleight:2017pcz} and discuss various related subtleties. 
Next, in section \ref{Mell} we propose a way to resolve a previously encountered problem with the definition
of the Mellin amplitude for the boundary correlator. Finally, we conclude in section \ref{conc}.

\section{Definitions of locality}
\label{loc}
First, let us specify what we mean by locality in AdS. For any scattering process in AdS one can evaluate the Witten
diagram, which results in some function ${\cal A}_n(x_i)$ of n points on the boundary $x_i$ associated with the external
lines of the scattering process. A particularly insightful representation for such amplitudes was recently proposed by Mack  
\cite{Mack:2009mi,Mack:2009gy}
\begin{equation}
\label{8sep1}
{\cal A}_n(x_i)=\prod_{1\le i<j\le n}\left( \int_{-i\infty}^{i\infty} \frac{d \delta_{ij}}{2\pi i}  \Gamma(\delta_{ij}) (x_{ij})^{-2\delta_{ij}} \right)
\prod_{i=1}^n \delta\left(\Delta_i - \sum_{j=1}^n \delta_{ij}\right)  {\cal M}_n(\delta_{ij}),
\end{equation}
where $x^2_{ij}=(x_i-x_j)^2$,
 $\Delta_i$ are dimensions of operators on external lines, $\delta_{ij}$ are variables dual to $x_{ij}$ and  ${\cal M}(\delta_{ij})$
is called the \emph{Mellin amplitude}. The integration contours for independent $\delta_{ij}$ that remain after solving the 
delta-function constraints in (\ref{8sep1}) run parallel to the imaginary axes 
in a way that the series of poles produced by each Gamma-function 
as well as by $M(\delta_{ij})$ stay on one side of the contours.

 The Mellin variables $\delta_{ij}$ can be thought of as the AdS counterparts of $q_i\cdot q_j$ 
for flat space scattering amplitudes, while operator dimensions $\Delta_i$ can be regarded as analogues of $m_i^2$.
Then, one can see that the delta-functions in (\ref{8sep1}) impose constraints on $\delta_{ij}$, which are equivalent to putting
external momenta on-shell and imposing momentum conservation in flat space. 
Along these lines, the function ${\cal M}(\delta_{ij})$ plays the role of the AdS counterpart of the flat scattering amplitude. 
Moreover, it was shown on numerous examples that Mellin amplitudes for scattering processes in AdS have a clear 
analytic structure similar to the analytic structure of flat space scattering amplitudes for the same processes
 \cite{Penedones:2010ue,Paulos:2011ie,Fitzpatrick:2011ia}. In particular, it was shown that contact interactions with a finite
 number of derivatives lead to polynomial Mellin amplitudes, while Mellin amplitudes for AdS exchanges are meromorphic
 functions featuring poles at locations, associated with dimensions of exchanged operators and their descendants.
We refer the reader to \cite{Penedones:2010ue,Paulos:2011ie,Fitzpatrick:2011ia} for more details.

This explains utility of the Mellin representation for studying locality in AdS. In flat space the theory is usually called local
if its Lagrangian has a finite number of derivatives. This implies that the amplitudes associated with contact interactions
should be polynomial. One can also consider a weaker notion of locality, which only demands that contact interactions 
produce amplitudes being entire functions. In this form, by employing Mellin amplitudes the notion of locality can 
be easily transferred from flat space to AdS. This is how (weak) locality was defined in \cite{Bekaert:2015tva} and we will adopt this definition
here.
To summarize, to verify whether the theory is local one would need to evaluate Mellin amplitudes for its contact diagrams and check
whether they are given by entire functions.

 First the issue of locality appears for quartic vertices and it is enough to consider self-interaction of scalar
fields. Using holography, the associated amplitude can be defined 
by subtracting contributions of four-point exchanges from the boundary correlator. Cubic vertices needed to define exchanges are determined 
by matching three point Witten diagrams with the associated three-point boundary correlators.
 This is the approach that was undertaken in  \cite{Bekaert:2014cea,Bekaert:2015tva}. 

Unfortunately, due to computational difficulties with spinning exchanges, completion of this program directly within the Mellin  representation remains
technically prohibiting. Instead, quartic self-interaction vertex was implicitly constructed in \cite{Bekaert:2015tva} using a certain spectral
representation for the conformal block decomposition, see  \cite{Sleight:2016hyl} for a comprehensive review on the topic.

Luckily, locality can also be translated into the language of  conformal blocks. 
In \cite{Heemskerk:2009pn} it was shown that all consistent four-point functions featuring only double trace conformal
blocks\footnote{Primary operators containing one/two trace contractions are called single/double trace operators. For vector models
these are bilinear/quadrilinear operators in elementary fields. Conformal blocks with single/double trace operators exchanged are called 
single/double trace conformal blocks.} in the conformal block decomposition are in one-to-one correspondence with contact four-point Witten diagrams in the bulk.
On the other hand, conformal block decomposition of exchanges in the direct channel contains  single trace conformal blocks
with dimensions equal to dimensions of  the exchanged operator. This suggests that absence/presence
of single trace conformal blocks can be used as an alternative criterion of locality/non-locality of the associated contact vertex, see \cite{Bekaert:2015tva}.

An important subtlety here is that even the exchange, while being non-local by definition,
 once expanded in the crossed channel, contains only double trace
conformal blocks in the conformal block decomposition \cite{Hoffmann:2000tr,ElShowk:2011ag}. This phenomenon is similar to the one for flat space amplitudes, when
the infinite series of local terms can hide a true singularity.
Let us note, however, that  in the same manner one can expect that an infinite series of single trace conformal
 blocks may, in principle, obscure the true locality nature of the vertex.

\section{Locality in holographic higher spin theory}
\label{lochs}

Let us now go into more details and see whether the quartic self-interaction vertex in the holographic higher spin theory
 is local by the single trace conformal block test.
 Below we will use the contact vertex in the crossing symmetric form, so it will be enough to check whether it has single
trace conformal blocks in the $s$-channel conformal block decomposition. As  explained above, the amplitude for the
contact four-point vertex is defined as
\begin{equation}
\label{8sep2}
{\cal A}_4^{c}(x_i) = G_4(x_i)-\sum_{l}\left( {\cal A}_{4}^{e|s,l}(x_i) + {\cal A}_{4}^{e|t,l}(x_i)+{\cal A}_{4}^{e|u,l}(x_i)\right),
\end{equation}
where $G\equiv \langle {\cal O}(x_1) {\cal O}(x_2){\cal O}(x_3) {\cal O}(x_4) \rangle_c$ is the connected part of the boundary correlator,
and ${\cal A}_{4}^{e|s,l}$, ${\cal A}_{4}^{e|t,l}$ and ${\cal A}_{4}^{e|u,l}$ are $s$-, $t$- and $u$-channel exchanges respectively 
with $l$ denoting spin of the exchange.

In \cite{Liu:1998th} it was shown that once cubic vertices in the bulk theory agree with the CFT side at  three-point level, 
then  exchanges accommodate all direct channel single trace conformal block contributions of the boundary correlator. In other
words, $s$-channel exchanges in (\ref{8sep2}) cancel all $s$-channel single trace contributions from the boundary correlator.
On the other hand, as it was already mentioned, each of the exchanges in the crossed channels has only double trace
conformal blocks in the $s$-channel conformal block decomposition. So, naively, one can conclude that the amplitude (\ref{8sep2})
for the contact vertex 
does not contain $s$-channel single trace conformal blocks and hence is local.

However, there is a flaw in this argument related to convergence of the spin sum for $t$- and $u$-exchanges.
Unfortunately, the $s$-channel conformal block decomposition for this sum cannot be evaluated explicitly. 
However, one can see that it may hide $s$-channel 
singularities as follows. 

First, we need to understand better the details of the conformal block decomposition of 
the boundary correlator. It reads
\begin{equation}
\label{8sep3}
\langle {\cal O}(x_1) {\cal O}(x_2){\cal O}(x_3) {\cal O}(x_4) \rangle_c = \frac{4}{N} \frac{1}{\left(x_{12}^2 x_{34}^2\right)^{\Delta}}
\left[ u^{\frac{\Delta}{2}}+ \left(\frac{u}{v} \right)^{\frac{\Delta}{2}}+ u^{\frac{\Delta}{2}} \left(\frac{u}{v} \right)^{\frac{\Delta}{2}}\right],
\end{equation}
where $u$ and $v$ are conformally invariant cross-ratios
\begin{equation}
\label{8sep4}
u = \frac{x_{12}^2 x_{34}^2}{x_{13}^2 x_{24}^2}, \qquad v = \frac{x_{14}^2 x_{23}^2}{x_{13}^2 x_{24}^2}
\end{equation}
and $\Delta\equiv d-2$ is the  dimension of the operator ${\cal O}=\phi^2$ of the free $O(N)$ vector model in $d$ dimensions.
For brevity, we denote the three terms on the right hand side of (\ref{8sep3}) as A, B and C.

It can be shown that ${\rm A}+{\rm B}$ contains only single trace conformal blocks in the $s$-channel conformal block decomposition, while
C contains only double trace blocks in the same channel.
 By doing cyclic permutations one can find analogous statements for conformal block decompositions in other channels.

Employing this information and the aforementioned result from \cite{Liu:1998th},
 we can conclude that the $t$-channel single trace contributions from ${\cal A}_{4}^{e|t,l}$ add up to ${\rm B}+{\rm C}$,
while the $u$-channel single trace contributions from ${\cal A}_{4}^{e|u,l}$ add up to ${\rm A}+{\rm C}$. In total this gives 
${\rm A}+{\rm B}+2{\rm C}$, which, besides a double trace contribution $2{\rm C}$,
contains a single trace piece ${\rm A}+{\rm B}$, when viewed from the point of view of the $s$-channel conformal block decomposition. This
indirect argument allows to show that, in fact, the contact vertex ${\cal A}_4^{c}(x_i)$ has the $s$-channel singularity equal to
minus that of the boundary correlator $G_4(x_i)$. Hence, the holographic higher spin theory is non-local 
by the single trace conformal block test, see \cite{Sleight:2017pcz}.

Let us point out few subtleties related to this argument. First of all, as we already mentioned, while presence of a finite series of
single trace conformal blocks in the conformal block decomposition does imply presence of poles in the Mellin amplitude and, therefore,
non-locality, it is not clear what kind of singularity, if any, may be associated with an infinite series of single trace conformal blocks.
For the case in question, the conformal block decomposition does contain an infinite series of single trace conformal blocks.
Though, it is hard to expect that the singularity is absent at all, it would be interesting to have a more qualitative understanding of what kind
of singularity we are dealing with. This is important at least to check whether this singularity is general enough to trivialise
the deformation procedure along the lines of the argument in \cite{Barnich:1993vg}.

Secondly, the argument given above, strictly speaking, applies to the common domain of validity of the conformal block
decompositions in all three channels, which is empty
\begin{equation}
\label{9sep1}
(u<1)\; \cap \; (v/u<1)\; \cap\; (v^{-1}<1) =   \varnothing .
\end{equation}
It was shown  that the domains of convergence of conformal block
decompositions are, in fact, much larger and can be applied to  correlators analytically continued by these decompositions
 \cite{Pappadopulo:2012jk,Rychkov:2015lca}.  While we do not expect any difficulties with the analytic
 continuation of (\ref{8sep3}) in the coordinate space, 
 for our purposes we rather need to make analytic continuations in the Mellin space, where neither
 we are aware of similar  convergence theorems  nor analytic continuations are straightforward
 if Mellin amplitudes involve distributions.

Finally, another subtlety is that exchanges besides single trace conformal blocks also inevitably contain double trace conformal
blocks in the conformal block decomposition. This double trace contribution can be altered by field redefinitions, or, equivalently,
by on-shell trivial cubic vertices. Nevertheless, it is not at all arbitrary. Singularities potentially can be generated from 
summation of these contributions over spin in the same way as it happens for single trace conformal blocks. This and some
other subtleties were discussed in \cite{Sleight:2017pcz}.

It is also interesting to confront the locality issue discussed here with its p-adic version. Holographic reconstruction of a 
quartic vertex in the p-adic case was performed in \cite{Gubser:2017tsi}. The striking difference with the Archimedean,
that is the standard,
analysis is that  due to peculiar properties of the p-adics, one does not have any spinning operators and
the sum (\ref{8sep2}) reduces to a single term with scalar exchanges. For this reason single trace contributions cancel
out on both sides and the quartic vertex is local.

\section{Mellin amplitude for the boundary correlator}
\label{Mell}

From the discussion in section \ref{loc} it is clear that the conclusion about locality in the holographic higher spin theory
depends exclusively on what the analytic structure of the Mellin amplitude for the boundary correlator is.
Let us clarify  what this amplitude is.

At four points (\ref{8sep1}) reads
\begin{align}
\notag
&{\cal A}_4(x_i)= \left(\frac{v}{x_{12}^2 x_{34}^2 }\right)^{\Delta}
\int_{c_s-i\infty}^{c_s+i\infty} \frac{ds}{2\pi i} \int_{c_t-i\infty}^{c_t+i\infty} \frac{dt}{2\pi i}\\
\label{11sep1}
& \qquad \qquad \qquad \qquad \cdot 
u^{s/2}v^{-{(s+t)}/{2}}
{\cal M}_4(s,t) \Gamma^2 \left[\frac{2\Delta-s}{2} \right]
\Gamma^2 \left[\frac{2\Delta-t}{2} \right] \Gamma^2 \left[\frac{2\Delta-u}{2} \right],
\end{align}
where we denoted 
\begin{align}
\notag
s\equiv \Delta_1+\Delta_2-2\delta_{12}= \Delta_3+\Delta_4-2\delta_{34},\\
\notag
t\equiv \Delta_1+\Delta_3-2\delta_{13}= \Delta_2+\Delta_4-2\delta_{24},\\
u\equiv \Delta_1+\Delta_4-2\delta_{14}= \Delta_2+\Delta_3-2\delta_{23},
\label{11sep2}
\end{align}
and then set $\Delta_i = \Delta$.
The Mandelstam variables $s$, $t$ and $u$ are analogous to those in flat space and satisfy
\begin{equation}
\label{11sep4}
s+t+u=4\Delta.
\end{equation}

The reason why the amplitude ${\cal M}$ is called the Mellin amplitude is because of its connection
to the Mellin transform of ${\cal A}$. The \emph{Mellin transform}  of a function $f(u)$ is defined by
\begin{equation}
\label{11sep5}
F(s)\equiv M[f(u)](s)\equiv \int_0^\infty du  f(u) u^{s-1}.
\end{equation}
This integral typically converges when $s$ belongs to a strip in the complex plane defined by $a<{\rm Re}(s)<b$ with $a$ and $b$ real.
 The Mellin transform  $F(s)$ is then analytic in this strip and this strip  is called the \emph{strip of analyticity} or \emph{the analyticity domain}. The inverse transform is given by
\begin{equation}
\label{11sep6}
f(u)=\int_{c-i\infty}^{c+i\infty}  \frac{ds}{2\pi i} F(s) u^{-s},
\end{equation}
where $a<{\rm Re}(c)<b$. In other words, the integration contour in (\ref{11sep6}) runs parallel to the imaginary axis anywhere within the strip
of analyticity.

By comparing (\ref{11sep1}) with the inverse Mellin transform formula (\ref{11sep6}), we can expect that the Mellin amplitude ${\cal M}$ can be obtained
from  the space-time amplitude ${\cal A}$ in the following two steps. First, one performs the Mellin transform of the amplitude ${\cal A}(u,v)$
expressed as a function of two independent cross-ratios $u$ and $v$ to find $M(s,t)$, called 
the \emph{reduced Mellin amplitude}
\begin{align}
{\cal A}_4(x_i)= \left(\frac{v}{x_{12}^2 x_{34}^2 }\right)^{\Delta}
\int_{-i\infty}^{i\infty} \frac{ds}{2\pi i} \int_{-i\infty}^{i\infty} \frac{dt}{2\pi i}
\label{11sep007}
u^{s/2}v^{-(s+t)/{2}}
M_4(s,t).
\end{align}
Next one finds the Mellin amplitude ${\cal M}_4(s,t)$ by factoring out the combination of Gamma-functions from 
the reduced Mellin amplitude $M_4(s,t)$
\begin{equation}
\label{11sep07}
M_4(s,t) = {\cal M}_4(s,t) \Gamma^2 \left[\frac{2\Delta-s}{2} \right]
\Gamma^2 \left[\frac{2\Delta-t}{2} \right] \Gamma^2 \left[\frac{2\Delta-u}{2} \right].
\end{equation}
However, as it is not hard to see, for the correlator (\ref{8sep3}) this procedure leads to the ill-defined 
Mellin amplitude.
 
Indeed, first problem that one encounters is the necessity to make the Mellin transform of the power function, which
according to the definition (\ref{11sep5})
 leads to 
an integral that diverges for any $s$. Still, there is a consistent framework that allows to define it as a 
distribution, see \cite{Bertrand:2000}
\begin{equation}
\label{11sep7}
   M[u^\Delta](s) \equiv \int_{0}^{\infty}  du \; u^{\Delta} u^{s-1} = \delta(\Delta+s).
\end{equation}
Note that here the strip of analyticity is understood as consisting of a single line ${\rm Re}(s)=\Delta$.
Then it is easy to verify that the inverse formula (\ref{11sep6}) does hold and the integration contour passes
right through the singularity of a delta-function.

Applying this formula to (\ref{8sep3}) we find
\begin{equation}
\label{11sep8}
M_4(s,t)= \frac{16}{N}\big(\delta(s-\Delta)\delta(t-\Delta)+\delta(s-\Delta)\delta(t-2\Delta)+\delta(s-2\Delta)\delta(t-\Delta)\big).
\end{equation}
Now we plug this into (\ref{11sep07}) to find the associated Mellin amplitude.
The reduced Mellin amplitude $M_4$ has  support consisting of only three points and the same should be
true for the Mellin amplitude ${\cal M}_4$ itself. Moreover, as it is not hard to see, for any of these three points the product of Gamma-functions
in (\ref{11sep07}) is singular. Hence,  the Mellin amplitude is the sum of  terms of the form $x\cdot\delta (x)$, which is zero as a distribution. If we keep the inverse transform formula (\ref{11sep1}) intact, the vanishing Mellin amplitude implies that the correlator  also vanishes in the coordinate representation, which is not the case. 
 This problem was encountered in \cite{Taronna:2016ats,Bekaert:2016ezc}.
 
 To summarize, the problem with a formal application of the rule that defines Mellin amplitudes for the case of free CFT's is as follows. The Mellin transform for each of the three terms in (\ref{8sep3}) is 
 known, well-defined, invertible and given in terms of distributions. Performing the Mellin transform of the free correlator in the coordinate representation we find the associated reduced Mellin amplitude.
   As a  final step, we are instructed to divide it  by the combination of Gamma-functions as in (\ref{11sep07}). Usually, the reduced Mellin amplitude is a genuine function and this step does not cause any problems. However, for free CFT's the reduced Mellin amplitude is a distribution, which is, moreover, supported on points where the double trace Gamma-functions are singular. As a result, an algebraic operation of  division by the product of double trace Gamma-functions is not invertible. One way to phrase this is to say that the Mellin amplitude for the free CFT correlator (\ref{8sep3}) is ill-defined.
 
It is worth to point out that the reduced Mellin amplitude (\ref{11sep8}) is also, strictly speaking, ill-defined, but for a different reason. As it was specified below (\ref{11sep7}), the Mellin transform of the power function is defined in a strip of analyticity consisting of a single line in the complex plane. For the three terms in (\ref{11sep8}) the strips of analyticity are
\begin{equation}
\label{28nov1}
\{{\rm Re}(s)=\Delta, \; {\rm Re}(t)=\Delta\}, \quad \{{\rm Re}(s)=\Delta, \; {\rm Re}(t)=2\Delta\}, \quad \{{\rm Re}(s)=2\Delta, \; {\rm Re}(t)=\Delta\}.
\end{equation}
They do not overlap, so the reduced Mellin amplitude (\ref{11sep8}) has the domain, which is the empty set.


\subsection{Inverse Mellin transform vs superposition of amplitudes}

In the previous discussion the Mellin amplitude was primarily understood as an alternative  representation for the Witten diagrams 
and conformal correlators. In this respect, it was important that there is an unambiguous connection between amplitudes in the standard coordinate representation and in the Mellin form.  So far this connection was realized via (\ref{8sep1}) and a prescription for the contour given below it. Similarly, the inverse relation is also known and involves the Mellin transform (\ref{11sep5}) as discussed in the previous section. For many relevant cases this dictionary is well-defined and is sufficient to establish an unambiguous connection between the Mellin and the coordinate representations.  There are, however, cases, where a naive application of this dictionary leads to an ill-defined result. One of these cases we encountered in the previous section. Below we will try to answer the question of what should be our guiding principle for defining Mellin amplitudes if the standard dictionary with the coordinate representation breaks down and how the Mellin amplitudes can be extracted once the coordinate representation of the respective amplitudes is known.

As this guiding principle we suggest a natural requirement that the Mellin amplitude for a superposition of processes is the sum of  Mellin amplitudes for each individual process. This \emph{superposition property} is absolutely standard for probability amplitudes, so it is natural to require that Mellin amplitudes satisfy it as well. In particular, it holds for amplitudes in the standard coordinate representation.
As we will see below, this property is in tension with the standard dictionary that relates the Mellin and the coordinate representations. More precisely, as it is not hard to see, the superposition property requires that the transform given by (\ref{8sep1}) is linear. This is only true if integration contours for all Mellin  amplitudes for superposed processes coincide. As it will be illustrated below, the standard locations of the contours, as defined below (\ref{8sep1}), for different processes, relevant for the holographic higher spin theory, are incompatible with each other. This implies that the Mellin amplitude for a superposition of such processes can not be related in the standard way to the associated amplitude in the coordinate representation.
For these problematic cases we propose to \emph{define} the Mellin amplitude for a superposition of processes as a sum of constituent Mellin amplitudes  irrespectively of the contour location constraints associated with these amplitudes. This enables us to extend the standard definition of the Mellin amplitude to cases for which it was previously inapplicable.

Below we will consider some natural examples of interfering processes in the holographic higher spin theories, for which the Mellin amplitudes defined in a standard way require incompatible integration contours.
 We will also discuss additional convergence subtleties that occur when the superposition involves an infinite set of processes.
Our main goal is to understand how the Mellin amplitude defined  by linearity as specified above can be related to the amplitude in the 
coordinate representation.
 Once this is clarified, we will propose  the Mellin amplitude for the free theory correlator. 

To start, we consider an example of the contact Witten diagram, for which the Mellin amplitude is free of singularities.
As it was mentioned below (\ref{8sep1}), the integration contours should not break any of  infinite series of poles generated by 
 $\Gamma(\delta_{ij})$. For the four-point case (\ref{11sep1}), this implies that the contour for $s$ integration should go between the poles generated by $\Gamma^2[(2\Delta-s)/2]$ and $\Gamma^2[(2\Delta-u)/2]$, while the contour for $t$ integration should separate the poles of $\Gamma^2[(2\Delta-t)/2]$ and $\Gamma^2[(2\Delta-u)/2]$. To this end, one should
require that 
\begin{equation}
\label{31nov1}
c_s<2\Delta, \qquad c_t<2\Delta, \qquad c_u<2\Delta,
\end{equation}
where  $c_u\equiv 4\Delta-c_s-c_t$.

\begin{figure}
\begin{center}
\includegraphics[width=0.7\linewidth]{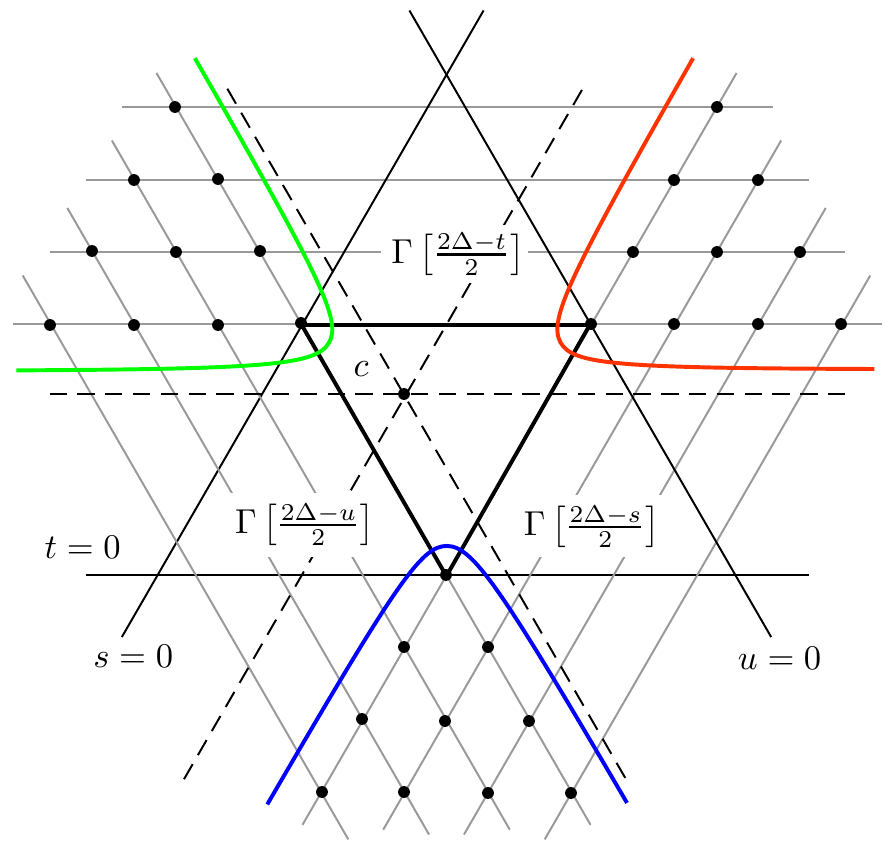}
\caption{This figure represents the real $(s,t,u)$ plane and locations of singularities of the reduced Mellin amplitude for the contact interaction. Singularities generated by $\Gamma(\delta_{ij})$ are shown as light-grey lines. The bold triangle is the real projection of the analyticity domain associated with the inverse Mellin transform of two variables, which defines admissible locations of the integration contour in the complex $(s,t,u)$ space. Depending on the way how we choose to close this integration contour at infinity, we pick one of the three sets of poles, enclosed in the red, the blue and the green contours.}
\label{fig:1}
\end{center}
\end{figure}

To reproduce the coordinate representation of the amplitude, we can first evaluate the $s$ integral in (\ref{11sep1}). Closing the contour at $s\to \infty$ and arguing that the infinite arc integral vanishes, we reduce this integration to the sum of residues  at $s=2\Delta+n$ with $n\ge 0$. This gives a power series in the cross-ratio $u/v$ with coefficients being functions of $t$. Then, for each term of the series we evaluate the $t$ integral. Closing the contour at $t\to \infty$ we pick residues at $t=2\Delta+m$ with $m\ge 0$, which produces an expansion in $v^{-1}$. Eventually, having evaluated both integrals, we find the amplitude presented as a power series in two cross-ratios $u/v$ and $v^{-1}$. The poles of the reduced Mellin amplitude, which residues were evaluated to arrive to this representation are enclosed inside the red contour on Fig. \ref{fig:1}.

Alternatively, one can close the integration contour at large $t$ and $u$, which produces an expansion of the amplitude in powers of the cross-ratios $u^{-1}$ and $v/u$. The associated poles are enclosed inside the green contour on Fig. \ref{fig:1}. Finally, by closing the contour at large $u$ and $s$ we produce an expansion in powers of $u$ and $v$. This expansion is generated by summing the residues of the reduced Mellin amplitude within the blue contour on Fig. \ref{fig:1}. The three expansions that we thus obtain are valid in the respective kinematics regimes. For domains where more than one expansion is valid, they produce the same result by virtue of various hypergeometric identities.

\begin{figure}
\begin{center}
\includegraphics[width=0.7\linewidth]{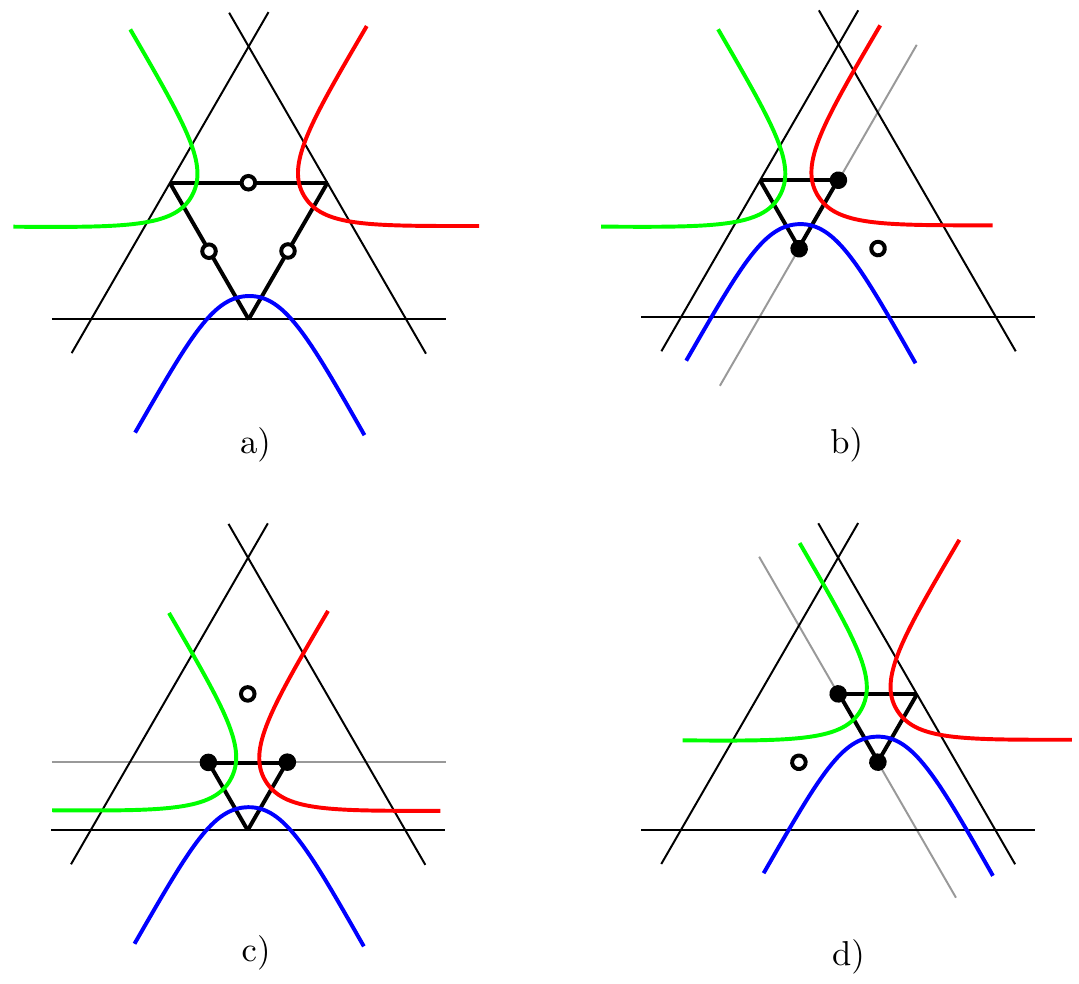}
\caption{Here we illustrate the singularity structures of the reduced Mellin amplitudes for the contact Witten diagram and for exchanges with the field of dimension $\Delta$ in $s$, $t$ and $u$ channels.
These are shown on figures $a)$, $b)$, $c)$ and $d)$ respectively. As before, bold triangles denote domains of analyticity of the reduced Mellin amplitude. Solid and empty circles represent locations of singularities associated with the three terms in the free theory correlator. Solid circles mean that the reduced Mellin amplitude has a given singularity, while empty circles mean that the reduced Mellin amplitude is regular at this point. The light-grey lines denote the leading single trace singularities of exchanges. For example, for the $s$-channel exchange it appears at $s=\Delta$.}
\label{fig:2}
\end{center}
\end{figure}

Let us now consider bulk exchanges with fields of dimension $\Delta$. The associated Mellin amplitudes are well-known \cite{Penedones:2010ue,Paulos:2011ie} and are given by hypergeometric functions. For generic dimensions of fields on external lines, the $s$-channel exchange Mellin amplitude features a series of poles at $s=\Delta+2n$ with $n\ge 0$. The requirement that these poles are not separated by the contour puts an additional constraint $c_s<\Delta$ on its location. So, combining all constraints together, one  finds that the domain of analyticity compared to the contact interaction case shrinks to, see Fig. \ref{fig:2},
\begin{equation}
\label{3nov1}
c_s<\Delta, \qquad c_t<2\Delta, \qquad c_u<2\Delta.
\end{equation}
As in the case of a contact interaction, there are three different ways to close the integration contour of the Mellin integral, which results in three alternative representations of the Witten diagram as a series in the conformal cross-ratios.

A special attention should be payed to singularities appearing at 
\begin{equation}
\label{3nov2}
(s,t,u): \qquad (\Delta,\Delta,2\Delta), \quad (\Delta, 2\Delta, \Delta), \quad (2\Delta, \Delta, \Delta)
\end{equation}
as these produce contributions that remain in the complete boundary correlator (\ref{8sep3}). The reduced Mellin amplitude for
the $s$-channel exchange has singularities at two of these locations. One is generated by 
\begin{equation}
\label{3nov3}
\frac{1}{(s-\Delta)(t-2\Delta)}
\end{equation}
and  contributes to the representation with the Mel\-lin integration contour closed at $s,t\to \infty$ (the red contour) and does not contribute to the others. Another singularity is of the form
\begin{equation}
\label{3nov4}
\frac{1}{(s-\Delta)(u-2\Delta)}
\end{equation}
and its residue only contributes to the integral with the contour closed at $s, u\to \infty$ (the blue contour).
Similar structures of poles and contour locations is exhibited by reduced Mellin amplitudes for other exchanges, see Fig. \ref{fig:2}.

It is now instructive to consider what happens if we add up three exchanges together. As we have just discussed, each of them has a well-defined Mellin amplitude, which allows to reproduce the associated amplitude in the coordinate representation using the standard dictionary. However, as it is not hard to see, admissible locations of the contours for different exchanges are incompatible with each other. For example, the analyticity domain for the $t$-channel exchange is
\begin{equation}
\label{3nov5}
c_s<2\Delta, \qquad c_t<\Delta, \qquad c_u<2\Delta
\end{equation}
and it has an empty overlap with the analyticity domain for the $s$-channel exchange (\ref{3nov1}). This manifests itself in a way that the singularity (\ref{3nov4}) at $(s,t,u)=(\Delta,\Delta,2\Delta)$ for the $s$-channel exchange is inside the blue contour,  while the singularity 
\begin{equation}
\label{3nov6}
\frac{1}{(t-\Delta)(u-2\Delta)}
\end{equation}
of the $t$-channel exchange is also located at the same point, but should be strictly outside the blue contour, as it is inside the green one. We would like to emphasize that contributions associated with these singularities are present in the free theory correlator, which means that  we should expect similar inconsistencies with the contours for its Mellin amplitude as well.

How should we proceed in this situation? As we discussed previously, it seems natural to define the Mellin amplitude for the process that involves three exchanges in different channels as a sum of individual Mellin amplitudes associated with each exchange. This is what we are expected to do for consistency with the amplitude's superposition principle. In this case, however, the standard relation between the Mellin and the coordinate representations breaks down. Instead, to reconstruct the coordinate representation of the amplitude from its Mellin form one should first split the amplitude into pieces and then use different contours for each of them. Of course, amplitudes can be split into parts in different ways, which will result in different outcomes in the coordinate representation. 
This means that under some circumstances \emph{Mellin amplitudes} do not define   amplitudes in the usual coordinate form
unambiguously, and, hence, \emph{do not give a faithful representation} for  scattering processes in AdS or, equivalently, correlators in the CFT.

It is also instructive to understand what happens with  integration contours when we sum an infinite series of Witten diagrams. For example, one can expand the $s$-channel exchange in a series of contact diagrams and consider Mellin amplitudes for each of these diagrams. Then, as it is not hard to see from Fig. \ref{fig:2}, the constraint for the location of the contour for these diagrams is different from that for the $s$-channel exchange. For example, the blue contour for contact interactions can go either below or above the point $(s,t,u)=(\Delta,\Delta, 2\Delta)$. At the same time, for the $s$-channel exchange diagram, the only prescription that gives the desirable result is when the blue contour goes above this point. The reason why this happens is clear. Namely, additional poles in the exchange Mellin amplitude imply that the sum over contact Mellin amplitudes does not converge everywhere. 
As this expansion is closely related to the Taylor expansion in $s$ at $s=0$, one would rather expect that it converges when the blue contour goes above the pole at $(s,t,u)=(\Delta,\Delta, 2\Delta)$, because then it is closer to $s=0$. From this example we learned that when dealing with series of Mellin amplitudes we may generate additional singularities and then the location of the contour should be chosen depending on where this series converges.

Another subtlety is related to the procedure that reduces integrals along the contours that go parallel to the imaginary axis to the sum of residues. For this procedure to commute with infinite summation, one has to ensure that the sum converges both in the strip of analyticity, where the initial contour runs, on the infinite arc contour that closes the initial contour and, in fact, everywhere inside the closed contour formed by joining these two contours together. To avoid these requirements one can deform the contour in the usual way before performing the summation. In this case it is enough to require that the series of Mellin amplitudes converges only in the vicinity of the real axes, where all their singularities are located. This, effectively, means that for the circumstances just described, one may reconstruct the Mellin amplitude from the coordinate representation by requirement that the reduced Mellin amplitude has correct residues irrespectively of what the standard Mellin integral gives.

\subsection{Regularized Mellin amplitude}

Previous considerations motivate the following modification of the standard dictionary between the Mellin and the coordinate forms of the free correlator. First of all, we leave the possibility of using different integration contours for different terms in the reduced  Mellin amplitude. To understand which contour should be chosen for each term, we will rely on how their singularities are located with respect to the contour in the reduced Mellin amplitudes for constituent Witten diagrams. Secondly, we replace the standard contour that runs parallel to the imaginary axis by a deformed one, which, essentially, replaces integration with the sum of residues of the reduced Mellin amplitude. As the standard integration contour can be deformed in different ways, we will require that sums of residues within each of the deformed contours produce the required coordinate representation.

With this said, let us adjust the reduced Mellin amplitude so that via the modified dictionary it translates into the free correlator (\ref{8sep3}) in the coordinate representation. First, we fix singularities located within the contour, which is closed at large values of $s$ and $t$, that is the red contour, see Fig. \ref{fig:1} and Fig. \ref{fig:2}.  
It already encloses singularities of the form
\begin{equation}
\label{4nov1}
\frac{1}{(s-\Delta)(t-2\Delta)}, \qquad \frac{1}{(s-2\Delta)(t-\Delta)},
\end{equation}
contributed by exchanges. Residues of these terms produce two out of three necessary contributions to the free theory correlator.
As it is not hard to see, for constituent amplitudes the red contour never encloses singularities capable to produce the remaining term. Still, considering that this term is present in the correlator,  the reduced Mellin amplitude should contain a singularity
\begin{equation}
\label{4nov2}
\frac{1}{(s-\Delta)(t-\Delta)}.
\end{equation}
As we discussed previously, this can happen as a result of summation of infinite series, e.g. exchanges over spin. Adding all contributions together, we find that the reduced Mellin amplitude, which is necessary to produce the boundary correlator from residues inside the red contour, reads\footnote{An analogous proposal for the Mellin transform for the power law function appeared in the context of the conformal bootstrap in Mellin space  \cite{Gopakumar:2016wkt}. }
\begin{equation}
\label{4nov3}
M_4^{r}=\frac{16}{N}\left( \frac{1}{(s-\Delta)(t-2\Delta)}+ \frac{1}{(s-2\Delta)(t-\Delta)}+\frac{1}{(s-\Delta)(t-\Delta)}  \right).
\end{equation}

Analogously, for the green and the blue contours we find
\begin{align}
\notag
M_4^{g}\;&=\frac{16}{N}\left( \frac{1}{(t-\Delta)(u-2\Delta)}+ \frac{1}{(t-2\Delta)(u-\Delta)}+\frac{1}{(t-\Delta)(u-\Delta)}  \right),\\
M_4^{b}\;&=\frac{16}{N}\left( \frac{1}{(u-\Delta)(s-2\Delta)}+ \frac{1}{(u-2\Delta)(s-\Delta)}+\frac{1}{(u-\Delta)(s-\Delta)}  \right).
\label{4nov4}
\end{align}
Then, the total reduced Mellin amplitude is given by
\begin{equation}
\label{4nov5}
M_4^w \equiv M_4^{r}+ M_4^{g}+M_4^{b}=0.
\end{equation}
Accordingly, the Mellin amplitude for the free theory correlator (\ref{8sep3})  is 
\begin{equation}
\label{4nov7}
{\cal M}_4^{w} = \frac{M_4^{r}+ M_4^{g}+M_4^{b}}{\Gamma^2\left[\frac{2\Delta-s}{2} \right]\Gamma^2\left[\frac{2\Delta-t}{2} \right]\Gamma^2\left[\frac{2\Delta-u}{2} \right]}
\end{equation}
and also \emph{formally} vanishes. Here ``formally'' additionally highlights the fact that cancellation occurs between Mellin amplitudes associated with incompatible integration contours.

In other words, we found that the total reduced Mellin amplitude for the free theory correlator is zero. Let us stress again, that this is not in contradiction with  the correlator being  non-zero in the coordinate representation --- this happens, because
 to recover the coordinate representation of the amplitude,  one has to use different integration contours for different terms in the reduced Mellin amplitude. To each term of the reduced Mellin amplitude (\ref{4nov5}) one can formally assign the following constraints on locations of  integration contours
\begin{align}
\notag
M^r: &\qquad c_s<\Delta, \qquad c_t<\Delta,\\
\notag
M^g: &\qquad c_t<\Delta, \qquad c_u<\Delta,\\
\label{4nov6}
M^b: &\qquad c_u<\Delta, \qquad c_s<\Delta.
\end{align}
However, let us emphasize again, that to recover the correlator in the coordinate representation, each of reduced Mellin amplitudes $M^r$, $M^g$ and $M^b$ should be integrated not along the standard contour that runs parallel to the imaginary axis, but rather along the deformed one, which runs in the vicinity of the real axis and  encloses all singularities of the reduced Mellin amplitude located there.

One may try to avoid a seemingly unattractive feature of this proposal of not having a single integration contour for all components of the reduced Mellin amplitude by various regularizations. Focusing first on the three terms in (\ref{4nov3}), (\ref{4nov4}) with a singularity at $(s,t,u)=(\Delta,\Delta,2\Delta)$, one can infinitesimally change these contributions, so that locations of singularities generated by these terms shift one from another, thus developing a common domain of analyticity. Once this is done, the inverse Mellin transform can be performed by simply adding these three contributions and 
 integrating them over a single integration contour inserted inside the common analyticity domain. Similar procedures can be performed with singularities at other locations.
An example of such regularization was recently considered in \cite{Rastelli:2017udc}.

It is worth to note that these regularizations do change the total reduced Mellin amplitude. This can be illustrated by the following simple one-dimensional example
\begin{equation}
\label{24nov1}
0 = \frac{1}{x}-\frac{1}{x} \quad\to \quad \lim_{\epsilon\to+0} \left(\frac{1}{x-i\epsilon}-\frac{1}{x+i\epsilon}\right)=2\pi i \lim_{\epsilon\to+0} \left(\frac{1}{\pi}\frac{\epsilon}{x^2+\epsilon^2} \right) = 2\pi i \delta(x),
\end{equation}
where regularization replaces zero by a delta-function. By virtue of analogous manipulations one can replace the initial amplitude 
(\ref{4nov3})-(\ref{4nov3}) by the one we started from in (\ref{11sep8}). As it was discussed above, the amplitude (\ref{11sep8}) still requires to use three different integration contours for each of the three terms in it. Hence, the necessity to use incompatible integration contours for different terms in the reduced Mellin amplitude cannot be avoided.

To summarize, in (\ref{4nov7}) we proposed an expression for the Mellin amplitude in the free CFT, which is defined not accordingly to a formal application of the Mellin transform as in \cite{Taronna:2016ats,Bekaert:2016ezc}, but rather as a sum of Mellin amplitudes for  constituent bulk Witten diagrams, which are well-defined and  known. Having studied how the integration contours of the inverse Mellin transform are located for each of these diagrams, we found that the total Mellin amplitude can be reconstructed from the boundary correlator in the coordinate representation by a certain modification of the standard procedure, which is described above. In particular, the standard integrals appearing in the inverse Mellin transform with integration contours going parallel to the imaginary axis were replaced by deformed ones going parallel to the real axis and enclosing singularities of the reduced Mellin amplitude located there. By doing that our goal was not to define a generalized version of the Mellin transform, but rather to reconstruct the Mellin amplitude, defined as a sum of constituent Mellin amplitudes,  without actually evaluating this sum and imposing as little requirements on the convergence of this sum as possible. One can imagine scenarios where such a contour deformation is not necessary and by evaluating the standard Mellin integrals along imaginary axes one still reproduces the required power function in the coordinate representation\footnote{Instead of  (\ref{4nov3})-(\ref{4nov5}) one could split zero into parts, so that each of them has a vanishing arc integral at infinity, which, in turn, ensures that the standard Mellin integral along the imaginary axis can be replaced by the sum of residues. If integration contours for different terms are located so that all singularities except one stay on one side of the contours, then contributions from these singularities cancels out due to the fact that the total reduced Mellin amplitude is vanishing. Then, the only non-vanishing contribution to the amplitude in the coordinate representation is given in terms of residues of the singularity that for different terms appears on different sides of integration contours. Clearly, this contribution will be given by the power function. Some concrete examples of this mechanism at work can be found in \cite{Rastelli:2017udc}.}. 
It would be interesting to see what actually happens by evaluating the sum of the bulk Witten diagrams in the Mellin representation explicitly.

\section{Conclusions}
\label{conc}

In conclusion,  we briefly discuss what (\ref{4nov7}) implies for locality of the holographic higher spin theory. First of all, given that the singular part of the Mellin amplitude for the  quartic scalar self-interaction in  the higher spin theory differs from (\ref{4nov7}) just by the sign,
we find that according to the definition of locality based on the analytic structure of Mellin amplitudes, the holographic higher spin theory is formally local. In all examples considered so far, this definition of locality appeared to be a successful counterpart of flat space locality defined in a standard way. In particular, in both flat and AdS spaces local interactions with a limited number of derivatives result into polynomial amplitudes, while  amplitudes for processes that involve exchanges have singularities. In other words, we believe that our formal conclusion about locality of holographic higher spin theories has good reasons to be taken seriously.

On the other hand, it appears that Mellin amplitudes do not represent faithfully all scattering processes in AdS, in a sense that different processes may have identical Mellin amplitudes. Moreover, this effect turns out to be crucial for amplitudes in the holographic higher spin theories. In particular, the Mellin amplitude for the conformal four-point correlator and, hence, the complete four-point higher spin scattering amplitude is formally zero. To be able to reconstruct a non-zero correlator in the coordinate representation, the Mellin amplitude should be first split into three parts and for each part one should use a different integration contour. The same effect is responsible for the formal vanishing of the singularity in the contact four-point interaction in the bulk higher spin theory. For this reason, one may argue that to make judgements about locality of the holographic higher spin theory, each of the three terms should rather be considered separately. As these terms contain singularities, one concludes that the associated non-localities are present in the bulk theory. In this way, our analysis can be reconciled with the conclusion of  \cite{Sleight:2017pcz} that the holographic higher spin theory is non-local.

At the same time, we would not go as far as claiming that presence of the singularity  (\ref{4nov7})
 trivializes the Noether procedure as a tool to construct
higher spin theories along the lines of the argument \cite{Barnich:1993vg}. What trivialises the Noether procedure is
generic non-localities associated with exchanges of fields present in the spectrum of the theory. For the holographic higher spin
theory, a generic $s$-channel exchange has the Mellin amplitude featuring a sequence of singularities at $s=\Delta, \Delta+2,\dots$. By inspecting explicit expressions in 
\cite{Penedones:2010ue,Paulos:2011ie} one finds that for generic space-time dimensions  this sequence is infinite.
On the contrary, leaving aside that  (\ref{4nov7}) formally vanishes, each of its terms
 has only a single singularity in each channel. In this respect it is more reminiscent of
exchanges with singletons on the boundary \cite{Paulos:2012nu,Nandan:2013ip,Nizami:2016jgt}, but with the dimension being
$\Delta$ (or $2\Delta$), not $\Delta/2$.
In fact, this similarity is not surprising given that the same singularity is produced by the exchange with an infinite tower of higher spin fields, which, in turn, via the Flato-Fronsdal theorem \cite{Flato:1978qz} can be related to a two-particle state of boundary singletons.

 It would be interesting to turn these observations into more concrete proposals of how the functional class of admissible Lagrangians
should be changed to make the Noether procedure for higher spin theories non-trivial. For example, one can propose that vertices that result into Mellin amplitudes with a finite number of poles are admissible\footnote{Four-point exchange diagrams with specially tuned dimensions of the exchanged field and fields on external lines may have a finite sequence of poles in the Mellin amplitude. This happens when the sequence of single trace poles of the reduced Mellin amplitude, e.g., $s=\Delta+2n$ with $n\ge 0$ overlaps with poles of double trace Gamma-functions. In this case all except a finite number of single trace singularities are cancelled in the Mellin amplitude by zeros from $\Gamma^{-1}(\delta_{ij})$. Of course, generic exchanges with fields in the spectrum of the theory should not be regarded as local interactions. This means that for the special values of dimensions of fields as discussed above, locality cannot be defined as a requirement that contact interactions result into Mellin amplitudes with finite sequences of poles.}.
Or, taking into account that the singularity of the four-point contact interaction formally vanishes in Mellin space, one may stick to the initial proposal of \cite{Bekaert:2015tva} and define locality as the requirement that the Mellin amplitude is an entire function.
 These proposals are based on a rather formal mathematical way
to describe  the difference between the known result (\ref{4nov7}) and a generic bulk exchange.
 Instead, it would be much more satisfactory if we had better understanding
of how different types of singularities in Mellin amplitudes manifest themselves in bulk experiments.
Then we would be able to motivate restrictions on functional classes of Lagrangians by the requirement
that these restrictions rule out only  theories that result into undesirable physical behavior. In this regard, let us stress again
that being dual to physically healthy theories on the boundary, higher spin theories are not expected to have serious physical pathologies.

It is worth to note that the Noether procedure with locality defined as the requirement that Mellin amplitudes for contact interactions
are free of poles  has the following unattractive property: it treats a sum of the four-point exchanges with holographically fixed cubic couplings as a local quartic interaction. This means, in particular, that one can rescale all cubic couplings by a spin-independent prefactor and by an appropriate local change of the quartic vertex keep the total four-point Witten diagram intact. It is not hard to see that this procedure does not violate neither the consistency conditions of the Noether procedure nor  locality, which, in turn, implies that cubic couplings may be fixed only up to an overall common factor.  It is suggestive  that the same pattern persists to all orders and the Noether procedure allows to define the theory up to one overall coupling constant at each order. In other words, it does not allow to reproduce the holographic higher spin theory unambiguously.

At the same time, such ambiguity may be used to achieve locality in the sense of the single trace conformal block test. Indeed, we can use the extra freedom to change the cubic couplings in a way that in (\ref{8sep2}) single trace conformal blocks cancel exactly.
One can proceed in a similar manner to higher orders. A higher spin theory so defined is consistent in the sense of the bulk Noether procedure and local in the sense of the single trace conformal block test. Of course, via holography it is incompatible with the CFT consistency conditions, more precisely, with the OPE. It would be still interesting to see whether this inconsistency has manifestations in terms of the purely bulk physics.

One may argue that technical  difficulties with evaluation of Mellin amplitudes for generic vertices make it hard to apply the amended Noether procedure in practice. In this respect we would like to stress that our goal in this paper was not to give a practical recipe for derivation of higher spin theories in AdS, but rather to explore at a formal level potential ways to define these theories perturbatively, taking into account a particular type of singularity present in holographic higher spin theories. After all,  higher spin theories in AdS can already be reconstructed from holography and do not need another derivation.
One still may hope that such analysis may lead to a better understanding of how locality should be relaxed and the Noether procedure be deformed for a more tractable problem of perturbative construction of  higher spin theories in flat space.

In summary, as it was shown in  \cite{Sleight:2017pcz},  the holographic higher spin theory features non-local interactions. Possibly, the most lucid way to convey this statement without going into technical details is to say that the singularity of the contact four-point interaction is proportional to the singularity of the sum of exchanges in the  theory.  At the same time, this singularity takes a  special form, which manifests itself, in particular, in the fact, that it vanishes in the appropriately defined Mellin representation. This information may be naturally used to define a non-trivial Noether procedure.
 It may also be instructive to view such a form of the singularity and of the total correlator in the Mellin representation
  as a consequence of the higher spin 
symmetry.  In this respect, we would like to note that in  the conformal higher spin theory
the symmetry forces amplitudes to vanish everywhere 
except for points of zero measure in kinematic space \cite{Joung:2015eny,Beccaria:2016syk}. It is also worth to note that in \cite{Sleight:2016xqq} the authors
argued that already the Lorentz part of the higher spin symmetry implies that amplitudes in higher spin theories should be
trivial. This conclusion is consistent with (\ref{4nov7}), if we  add all contributions to the Mellin amplitude together. On the other hand, the correlator in the coordinate representation is non-vanishing and still covariant with respect to the higher spin symmetry. This once again highlights inequivalence between two languages - the Mellin and the coordinate representation -  and may hint towards a way to go around  no-go theorems for higher spin interactions in flat space.

\paragraph{Note added:} In the initial version of this paper we made a general observation that for constituent Witten diagrams the integration contour appearing in the Mellin transform encloses singularities, but does not go exactly through them as for the power law function. Based on that we suggested that the Mellin amplitude in the free CFT should be a rational function with simple poles at the required locations. The remaining ambiguities were fixed in a heuristic way. Then, in \cite{Rastelli:2017udc}  the four-point correlator in IIB supergravity was computed. It was found that in Mellin representation the free part of the correlator arises as a regularization effect in the inverse Mellin transformation. This lead us to extend our previous analysis by carefully taking into account where the contour is located for each constituent Witten diagram.
In agreement with  \cite{Rastelli:2017udc} we found, that there is a consistent sense in which the Mellin amplitude can be defined to be zero.

\section*{Acknowledgements}
 I am grateful to S. Sarkar,  E. Skvortsov and A. Tseytlin for fruitful discussions on different aspects of this work.
 I would also like to thank X. Bekaert,  E.~Skvortsov and A.~Tseytlin for comments on the draft. 
  This work was supported by the ERC Advanced grant No.290456.

\bibliography{KHSA}
\bibliographystyle{utphys}
\end{document}